\newcommand{\be}{\begin{equation}}
\newcommand{\ee}{\end{equation}}
\newcommand{\bea}{\begin{eqnarray}}
\newcommand{\eea}{\end{eqnarray}}
\newcommand{\sect}{\section}
\newcommand{\al}{\alpha}
\newcommand{\pa}{\partial}

\newcommand{\ga}{\gamma}
\newcommand{\vrho}{\varrho}

\newcommand{\ka}{\kappa}
\newcommand{\de}{\delta}

\newcommand{\vphi}{\varphi}
\newcommand{\th}{\theta}

\newcommand{\Ups}{\Upsilon}
\newcommand{\rar}{\rightarrow}

\newcommand{\pap}{\partial_{+}}
\newcommand{\non}{\nonumber}

\newcommand{\ppm}{{\rm P\!}_{\pm}}

\newcommand{\ra}{\rangle}
\newcommand{\la}{\langle}
\newcommand{\cG}{{\cal G}}
\newcommand{\cA}{{\cal A}}
\newcommand{\cW}{{\cal W}}

\newcommand{\cD}{{\cal D}}

\newcommand{\gn}{{\cal G}_{0}}
\newcommand{\gm}{{\cal G}_{-}}
\newcommand{\emt}{energy momentum tensor }
\newcommand{\ba}{\begin{array}}
\newcommand{\ea}{\end{array}}
\documentstyle[12pt,twoside]{article}
\include{psfig}
\begin{document}
\vspace{-1mm}
\begin{flushright} G\"{o}teborg ITP 96-6\phantom{i} \\ hep-th/9609125
\end{flushright}
\vspace{1mm}
\begin{center}{\bf\Large\sf Generalized  $N=2$ Supersymmetric Toda Field
Theories}
\end{center}
\vspace{1mm}
\begin{center}{{\bf\large Niclas Wyllard{\normalsize
\footnote{wyllard@fy.chalmers.se}}\vspace{5mm}}\\{\em Institute of Theoretical
Physics, S-412 96 G\"{o}teborg, Sweden}}
\end{center}
\begin{abstract}
In this paper we introduce a class of generalized supersymmetric Toda field
theories. The theories are labeled by a continuous parameter and have $N=2$
supersymmetry. They include previously known $N=2$ Toda theories as special
cases. Using the WZNW$\rightarrow$Toda reduction approach we obtain a closed
expression for the bracket of the associated $\cW$ algebras. We also derive an
expression for the generators of the $\cW$ algebra in a free superfield
realization.
\end{abstract}
\setcounter{equation}{0}
\sect{Introduction}
In a recent paper \cite{me} we considered a certain class of generalized
(classical) conformal Toda field theories, proposed earlier by Brink and
Vasiliev \cite{Brink}. It is natural to extend this construction to
supersymmetric Toda theories. Supersymmetric conformal Toda theories based on
finite-dimensional simple Lie superalgebras have been described previously in
the literature \cite{Evans-Hollowood,Saveliev,Olshanetsky}. In this paper we
will describe a class of infinite-dimensional generalized supersymmetric Toda
theories. The theories are labeled by a continuous parameter, such that when
this parameter takes certain discrete values the model reduces to a
finite-dimensional supersymmetric Toda theory.
The generalized models will turn out to have $N=2$ supersymmetry. Through a
reduction procedure, which will be described in more detail later, it is also
possible to obtain generalized Toda theories with $N=1$ supersymmetry. We will
only discuss the classical models, and leave the quantization for a future
publication.
The paper is organized as follows. In the next section we discuss the general
Lie superalgebra on which our construction is based, and derive some results
which will be used in later sections.
In section \ref{swznw} we define and describe the generalized models, using the
WZNW reduction approach to supersymmetric Toda theories, developed earlier for
the finite-dimensional cases \cite{Delduc}. Associated with the generalized
Toda models we find a set of generalized $\cW$ algebras.
As in the bosonic case \cite{me} it is possible to obtain a closed expression
for the bracket used to calculate the $\cW$ algebra associated with the
generalized Toda theories. The calculation of this (Dirac) bracket is presented
in section \ref{sdb}. Finally, in the last section we discuss a free superfield
realization of the general $\cW$ algebra, and give a closed expression for the
$\cW$ algebra generators in terms of these free fields, i.e. a Miura
transformation. We apply the method to a simple case as an example.
The (finite-dimensional) conformal supersymmetric Toda theories are described
in detail in \cite{Evans-Hollowood}. For more about WZNW$\rar$Toda reduction
the review \cite{ORPhysrep} contains useful information.

\setcounter{equation}{0}
\sect{The General Lie Superalgebra} \label{salg}
The general Lie superalgebra upon which our construction is based has been
described before \cite{Vasiliev89,Vasiliev91,Brink,me}. We will therefore
concentrate on the points which are most relevant for the present paper. The
general Lie superalgebra is defined as the algebra whose elements are monomials
in the three operators $K$ and $a^{\pm}$ satisfying
\bea
	[ a^{-},a^{+} ] &=& 1 + 2\nu K \,, \non \\ \{ K,a^{\pm}\} &=& 0 \;\; ,\;\;\;\;
 K^{2}=1\,.
\eea
A general element can therefore be written as
\be
B = \sum_{A=0}^1 \sum_{n=0}^\infty \frac{1}{n!}
b^A_{\, \alpha_1 \ldots \alpha_n } K^A\,
a^{\alpha_1} \ldots a^{\alpha_n }\,.\label{gel}
\ee
We can choose $b^A_{\, \alpha_1 \ldots \alpha_n }$ to be totally symmetric in
the lower indices, so the basis elements can be chosen to be the Weyl ordered
products
\be
\label{base}
E_{nm}=\Big ( (a^+ )^n (a^- )^m \Big )_{\rm Weyl} =\frac{1}{(n+m)!}
\Bigl ( (a^+ )^n (a^- )^m +((n+m)!-1)  )\,\mbox{permutations}\Bigr ) \,,
\ee
multiplied by an appropriate function of the Klein operator $K$. Because of the
Klein operator there are two independent basis elements for fixed $n$ and $m$.
The (general) algebra possesses a natural ${\bf Z}_{2}$ gradation. The grade
$[A]$ of an element $T^{A}$ in the algebra is defined to be equal to $0$ or
$1$, for monomials with even or odd powers $n$ in (\ref{gel}), respectively.
The graded bracket satisfy the usual property
\be
	 [T^{A},T^{B}\} = -(-1)^{[A][B]}[T^{B},T^{A}\}\,.
\ee
The operators $T^{\pm}=\frac{1}{2}(a^{+})^{2}$ and
$T^{0}=\frac{1}{4}\{a^{-},a^{\pm}\}$ span an $sl_{2}$ subalgebra of the general
Lie superalgebra. An important point for this paper is that the general algebra
has an $osp(1|2)$ subalgebra
\bea
	\{ a^{\pm},a^{\pm}\} = 4T^{\pm} &,&  \{ a^{+},a^{-}\} = 4T^{0}\,, \non \\
        {[} T^{0},a^{\pm}] = \pm\frac{1}{2}a^{\pm} &,& [T^{\pm},a^{\mp}]=\mp
a^{\pm}\,, \non \\
	{[}T^{0},T^{\pm}] = \pm T^{\pm} &,& [T^{-},T^{+}]=2T^{0}\,, \label{osp}
\eea
which can be extended to an $sl(2|1) \cong osp(2|2)$ subalgebra
\bea
	 \{ \bar{c}^{+},\bar{c}^{-}\} = 2h &,&  \{ c^{+},c^{-}\} = 2\bar{h} \,, \non
\\
        {[} h,c^{\pm}] = \pm c^{\pm}&,& [ \bar{h},\bar{c}^{\pm}] =
\pm\bar{c}^{\pm}\,, \non \\
        \{ c^{\pm},\bar{c}^{\pm}\} = 2T^{\pm} &,& [T^{\pm},c^{\mp}]=\mp
\bar{c}^{\pm} \,,\non \\
	{[}T^{\pm},\bar{c}^{\mp}]=\mp c^{\pm} &,& [h,T^{\pm}] = \pm T^{\pm} \,, \non
\\      {[}\bar{h},T^{\pm}] = \pm T^{\pm} &,&  [T^{-},T^{+}]=h + \bar{h}\,.
\eea
Here
\bea
	 c^{\pm} = \ppm a^{\pm} &,&  \bar{c}^{\mp} = \ppm a^{\mp}\,, \non \\
	 h = T^{0} + \frac{K}{4}(1 + 2\nu K) &,& \bar{h} = T^{0} - \frac{K}{4}(1 +
2\nu K)\,,
\eea
and $\ppm = \frac{1 \pm K}{2}$.
The algebra has a unique invariant supertrace operation
\cite{Vasiliev89,Vasiliev91}, defined as
\be
\label{str}
str (B) = b^0 -2\nu b^1 \,,
\ee
where $B$ is of the form (\ref{gel}), with the property
\be
\label{strprop}
str (T^{A}T^{B}) = (-1)^{[A]}str (T^{B}T^{A}) = (-1)^{B}str(T^{B}T^{A}) \,.
\ee
The properties of the bilinear form $\la\cdot,\cdot\ra$ defined as $\la
T^{A},T^{B}\ra = str(T^{A}T^{B})$, can thus be collected together as
\be
	\begin{array}{rccl}
		\la T^{A},T^{B}\ra &=& 0, \;\;  {\rm if} \;\; [A] \neq [B] & {\rm
(consistency)}\\
		\la T^{A},T^{B}\ra &=& (-1)^{[A][B]}\la T^{B},T^{A}\ra & {\rm
(supersymmetry)} \\
		\la [T^{A},T^{B}\}T^{C}\ra &=& \la T^{A}[T^{B},T^{C}\}\ra   & {\rm
(invariance)}\,.
	\end{array} \label{biprop}
\ee
We also have the formula
\be
	str(K^{A}E_{nm}E_{rs}) = \de_{mr}\de_{ns}(-1)^{n}n!m!\beta^{A}(n+m)
\label{EE}\,,
\ee
with $\beta^{A}(0) = [\de_{A,0} - 2\nu\de_{A,1}]$ and
\be
	\beta^{A}(n) = 2^{-n}\left[ \de_{A,0} -
\frac{1}{2}(1+(-1)^{n})\frac{2\nu}{n+1}\de_{A,1}\right]
\prod_{l=0}^{[(n-1)/2]}(1-\frac{4\nu^{2}}{(2l+1)^{2}})\,.
\ee
The algebra under discussion can be shown to be isomorphic to the universal
enveloping algebra of $osp(1|2)$, $U(osp(1|2))$, divided by a certain ideal
\cite{Bergshoeff91}.  When $d=n$, where $d=\frac{2\nu +1}{2}$ and $n$ is a
positive integer, the bilinear form (pseudo scalar product) $\la\cdot,\cdot\ra$
becomes degenerate i.e. ``null-states'' appear, and the algebra becomes, after
dividing out the ideal spanned by the null-states, isomorphic to $gl(n|n-1)$
when considered as a Lie superalgebra. {}From now on we will assume that the
ideal $1$ ($=E_{00}$) has been factored out from the general Lie superalgebra;
this means that we get $sl(n|n-1)$, when $d=n$ (after the ideal spanned by the
null-states has been divided out).
{}From the general algebra it is also possible to obtain the algebras
$B(n,n)\cong osp(2n-1|2n)$ and $B(n,n-1)\cong osp(2n+1|2n)$ through the
following procedure \cite{Vasiliev91}. First we introduce the
anti-automorphism\footnote{Recall that an anti-automorphism $\rho$ satisfies
$\rho(UV)=\rho(V)\rho(U)$, whereas an automorphism $\tau$ satisfies
$\tau(UV)=\tau(U)\tau(V)$.} $\rho(a^{\pm})=ia^{\pm}$ and $\rho(K)=K$.
{}From $\rho$ we can obtain an automorphism $\tau$ through $\tau(T^{A}) =
-i^{[A]}\rho(T^{A})$. Then by extracting the subset of elements satisfying
$\tau(T^{A})=T^{A}$ we get (after the appropriate ideal has been divided out)
$B(n,n)$ when $d=2n$ and $B(n,n-1)$ when $d=2n+1$. It is not possible to obtain
the other simple Lie superalgebras through this construction.

It will turn out to be convenient to use the following basis of the algebra.
For $n+m =$ even, we define
\bea
	E^{1}_{nm} &=& E_{nm} \,, \non \\
	E^{2}_{nm} &=& \frac{m+1}{2}(K + \frac{2\nu}{m+n+1})E_{nm}\,,
\eea
and for $n+m =$ odd
\bea
	E^{1}_{nm} &=& E_{nm} \,, \non \\
	E^{2}_{nm} &=& KE_{nm}\,.
\eea
{}From (\ref{EE}) it follows that the basis elements satisfy
\be
\ba{lcll}
	str(E^{1}_{nm}E^{1}_{mn}) &=& - str(E^{2}_{n-1,m-1}E^{2}_{m-1,n-1}) & (n+m
={\rm even}) \,, \non \\str(E^{1}_{nm}E^{1}_{mn}) &=& -
str(E^{2}_{n,m}E^{2}_{m,n})& (n+m={\rm odd}) \,, \non \\
	str(E^{1}_{nm}E^{2}_{pr}) &=& 0\,. &
\ea \label{orto}
\ee
We will call those elements which are annihilated by $ad_{a^{+}}$
($ad_{a^{-}}$)  highest (lowest) weight elements. We also define the following
operators
$T = ad_{T^{+}} = [T^{+},\cdot \}$, and $A = ad_{a^{+}}=[a^{+},\cdot\}$. The
action of $T$ on the basis elements is
\bea
	[T^{+},E_{n,m}\} = -mE_{n+1,m-1} &,& [T^{+},E_{n,0}\} = 0\,. \label{Tprop}
\eea
In later sections we will need the action of $A$ on the basis elements. We
have, when $n+m$ is even
\bea
	[a^{+},E^{1}_{n,m}] &=& -mE^{1}_{n,m-1}\,, \non \\
	{[}a^{+},E^{2}_{n,m}] &=& -(m+1)E^{2}_{n+1,m}(1-\de_{m,0}) \,,
\eea
and when $n+m$ is odd
\bea
	\{a^{+},E^{1}_{n,m}\} &=& 2E^{1}_{n+1,m} \non \,, \\
	\{a^{+},E^{2}_{n,m}\} &=& 2E^{2}_{n,m -1}(1-\de_{m,0})\,.
\eea
These relations are easily derived e.g. by starting from $E^{1}_{0,2n}$ and
$E^{2}_{0,2n+1}$, i.e. lowest weight elements, and using the identity $AT^{m} =
T^{m}A$. The action of $T$ on the basis elements is previously known (cf.
(\ref{Tprop}), and the action of $A$ on lowest weight elements is easy to
calculate. The properties above mean that $A$ maps the two sectors of the basis
elements into themselves (with the exception of highest weight elements, i.e.
those elements which are annihilated by $ad_{a^{+}}$.) The basis elements can
be arranged in the ``wedge'' shown on the next page.
\begin{figure}[t]
   \centerline{\psfig{figure=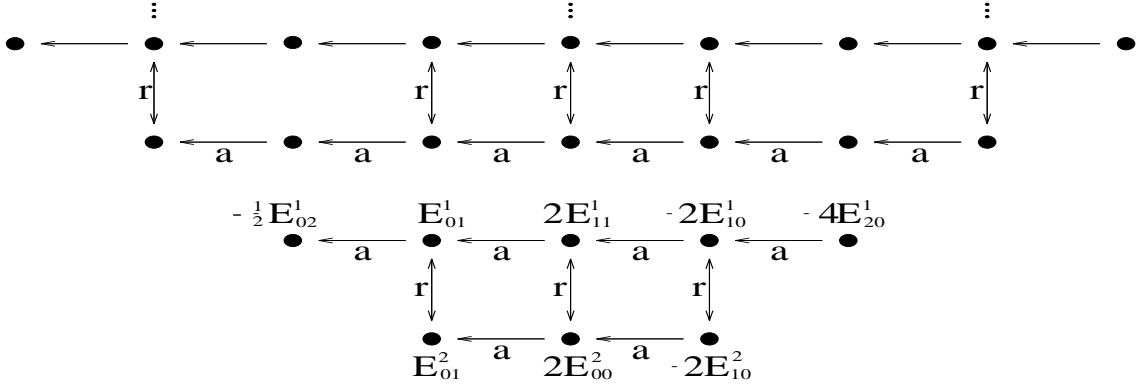,height=5cm,width=15cm}}
   \caption{The wedge of $osp(1|2)$ irreducible representations.}
\end{figure}
The horizontal lines are  irreducible representations of $osp(1|2)$, and $A$ is
the step-operator of these representations. This fact makes the orthogonality
of the basis elements (\ref{orto}) transparent. We will denote the ``inverse''
of $A$ by $a$. We have
\bea
	[a^{+},a(x)\} &=& x - \Pi_{-}(x) \,,\non \\
	a([a^{+},x\}) &=& x - \Pi_{+}(x) \,,
\eea
where $x$ is an arbitrary element in the algebra, $\Pi_{+}$ is the projection
operator which projects onto the space of highest weight elements, and
similarly $\Pi_{-}$ projects onto lowest weight elements. The action of $a$ on
the elements in the algebra, can be inferred from the action of $A$ described
above. Furthermore, the following properties are satisfied by $A$ and its
``inverse'' $a$
\bea
	\la UA(V)\ra = - \la A(\hat{U})V\ra &,& \la Ua(V)\ra = \la a(\hat{U})V\ra
\label{aprop}\,,
\eea
where $U$ and $V$ are arbitrary elements of the algebra. The operator
$\hat{\phantom{a}}$ acts in the following way $\hat{\phantom{a}}\; : \;T^{A}
\rightarrow (-1)^{[A]}T^{A}$ on the algebra elements, and is called the grading
automorphism.
We define the operator $r$ in the following way. When $m+n$ is even
\bea
	r(E^{i}_{nm}) = E^{i}_{nm}\;\;(n>m) &,& r(E^{i}_{nm}) = -E^{i}_{nm}\;\;(n<m)
\,, \non \\
       r(E^{2}_{2n,2n}) &=& E^{1}_{2n+1,2n+1}\,,\non \\
       r(E^{1}_{2n+1,2n+1}) &=& E^{2}_{2n,2n} \,,
\eea
and when $n+m$ is odd
\bea
	r(E^{1}_{nm}) &=& E^{2}_{nm} \non \,, \\
	r(E^{2}_{nm}) &=& E^{1}_{nm} \,.
\eea
It follows from these definitions that
\bea
	r^{2}=1 &,& \la r(U)V\ra = - \la Ur(V)\ra \label{alm}\,,
\eea
and
\be
	r([U,r(V)\}) + r([r(U),V\}) - [r(U),r(V)\} = [U,V\}\,. \label{yb}
\ee
The first relation is immediate. In the fermionic sector the second relation
follows from the fact that in this sector $r$ is equal to multiplication by
$K$, together with the properties of the supertrace; whereas in the bosonic
sector it follows from the properties of the supertrace. E.g. for the neutral
elements the following combinations
\bea
	h_{n} = \frac{E^{1}_{nn} + E^{2}_{n-1,n-1}}{2} &,& \bar{h}_{n} =
\frac{E^{1}_{nn} - E^{2}_{n-1,n-1}}{2}\,,
\eea
satisfy $r(h_{n}) = h_{n}$, $r(\bar{h}_{n}) = -\bar{h}_{n}$, and
\bea
	str(h_{n}h_{m}) =0 &,&  str(\bar{h}_{n}\bar{h}_{m}) = 0 \,,\non \\
	str(h_{n}\bar{h}_{m}) &\propto& \de_{nm} \,.
\eea
Equation (\ref{yb}) follows from the fact that $r$ is equal to $K$ when acting
on fermions, together with the fact that the commutator of e.g. two bosonic
elements with positive grade is another bosonic element with positive grade.
Equation (\ref{yb}) is the (classical) modified Yang-Baxter equation. Another
interpretation is the following. If we naively set $J=ir$, then (\ref{alm})
states that $J$ satisfies the properties of an almost complex structure which
is compatible with the metric, whereas (\ref{yb}) implies the vanishing of
\be
	N(U,V) = [U,V\} + J([U,J(V)\})+ J([J(U),V\}) - [J(U),J(V)\} \,, \label{nij}\ee
which we recognize as the Nijenhuis tensor (field). Hence $J$ is promoted to a
complex structure. However, we should remember that we are considering a real
algebra and the operator $J$ does not respect the reality condition.
Nevertheless, the existence of the operator $r$ is crucial for the appearance
of $N=2$ supersymmetry in the Toda theories described later, which are based on
the algebra in this section. However, the supersymmetry is not of the standard
form due a sign discrepancy between the algebra of the two sets of supercharges
(caused by the fact that $r$ squares to 1 and not to $-1$). For a discussion of
this point we refer to ref. \cite{Evans}.

\setcounter{equation}{0}
\sect{WZNW$\rar$Toda Reduction} \label{swznw}
In this section we will describe the generalized Toda theories using the
WZNW$\rar$Toda reduction approach.
We will start with a brief recapitulation of the procedure for the finite
dimensional cases treated earlier \cite{Delduc,Inami}. The number of
supersymmetries in the resulting Toda theories are related to the existence of
certain subalgebras of the underlying Lie superalgebra.
These so called principal subalgebras correspond to the different number of
supersymmetries which are possible, according to the following rule: An
$sl_{2}$ principal subalgebra gives rise to $N=0$, an $osp(1|2)$ principal
subalgebra leads to $N=1$ and finally the existence of an $sl(2|1)\cong
osp(2|2)$ principal subalgebra implies $N=2$ supersymmetry. The generalized
Toda theories studied in this paper have $N=2$ supersymmetry, thus generalizing
the finite-dimensional theories based on the $sl(n|n-1)$ Lie superalgebras.
We would like to keep the $N=2$ supersymmetry manifest during the discussion,
but unfortunately at the moment the $N=2$ WZNW model in $N=2$ superspace is not
known. We will therefore (almost exclusively) work in $N=1$ superspace. See
however ref. \cite{Magro}, which discusses certain aspects of a manifestly
$N=2$ treatment. In the $N=1$ formalism the $N=2$ symmetry is hidden.
The WZNW action in $N=1$ superspace is\footnote{Conventions: $D_{\pm} =
\frac{\pa}{\pa\th^{\pm}} + \th^{\pm}\pa_{\pm}$, $D^{2}_{\pm} = \pa_{\pm}$.}
\bea
	S_{W}(g)&=&\frac{\ka}{2}\int d^{2}z
d^{2}\th\la(G^{-1}D_{+}\hat{G}),(\hat{G}^{-1}D_{-}G)\ra \non\\ &+&
\frac{\ka}{2}\int dtd^{2}zd^{2}\th \la
G^{-1}\pa_{t}G,(G^{-1}D_{+}\hat{G}\hat{G}^{-1}D_{-}G +
G^{-1}D_{-}\hat{G}\hat{G}^{-1}D_{+}G )\ra \,. \label{wzwact}
\eea
The superfield $G(z,\th)$ takes its values in a supergroup, and is of the
general form
\be
	G(z,\th) = \exp(\Phi_{i}(z,\th)B^{i} + \Psi_{i}(z,\th)F^{i})\,.
\ee
Here $B^{i}$ belongs to the set of basis elements of the subalgebra with ${\bf
Z}_{2}$ grade 0, and $F^{i}$ has ${\bf Z}_{2}$ grade 1. $\Phi_{i}$ is a
Grassmann even superfield and $\Psi_{i}$ is Grassmann odd. By identifying the
two different gradations, we call a Lie superalgebra valued superfield odd if
its total grade is odd, and even if its total grade is even.
In this language the WZNW field $G$ is thus even. The grading automorphism
$\hat{\phantom{a}}$ was defined in section \ref{salg}. It has the following
action on $G$, $ \exp(\Phi_{i}B^{i} + \Psi_{i}F^{i})\rightarrow
\exp(\Phi_{i}B^{i} - \Psi_{i}F^{i})$. From now on we set $\ka =1$. The WZNW
currents obtained from the action (\ref{wzwact}) are
\bea
	J_{+} = \hat{G}^{-1}D_{+}G &,& J_{-} = (D_{-}G)G^{-1}\,.
\eea
These currents can be expanded as $J = \Psi_{i}B^{i} + \Phi_{i}F^{i}$, i.e.
they are odd in the terminology introduced above.
The equation of motion can be written as
\bea
	 D_{-}J_{+} = D_{-}(\hat{G}^{-1}D_{+}G) = 0 &\Leftrightarrow&   D_{+}J_{-} =
D_{+}( (D_{-}G)G^{-1}) = 0\,.
\eea
We will denote the Lie superalgebra obtained from the supergroup by $\cG$. The
algebra is assumed to have an $osp(1|2)$ principal subalgebra of the form
(\ref{osp}). Algebras of this type include the finite-dimensional algebras
listed in \cite{Evans-Hollowood} as well as the general algebra and its
truncations described in the previous section. In the sequel the underlying Lie
superalgebra $\cG$ is assumed to be the general algebra. In this case however,
all statements which involve the concept of a group should be taken to be true
on a formal level. The method given below is a formal way to obtain the
generalized Toda theories; these can then be defined by their action and
associated equations of motion.
To obtain the generalized Toda theories the following constraints are imposed
on the currents
\bea
	\la E_{\al^{+}},D_{-}GG^{-1} - a^{-}\ra &\approx& 0\,, \non \\
	\la E_{\al^{-}},\hat{G}^{-1}D_{+}G -a^{+}\ra &\approx& 0\,, \label{constr}
\eea
for all basis elements $E_{\al^{+}}$ with positive $T^{0}$ grade and all basis
elements $E_{\al^{-}}$ with negative grade,
which means that, on the constraint surface, the currents become
\be
	J_{\pm} = a^{\pm} + j_{\pm}\,.
\ee
Here $j_{\pm}$ lies in the Borel subalgebra $\gn\oplus{\cal G}_{\mp}$, where
$\cG_{\pm}$ are the parts of the algebra with positive and negative $T^{0}$
grade respectively, and $\gn$ is the subset of neutral elements. There is also
an alternative lagrangean realization of the reduction in terms of a gauged
WZNW model in $N=1$ superspace, but we will not use it here.
Continuing with the general scheme, we make the Gauss
decomposition\footnote{This decomposition is only valid locally. We do not
consider global issues \cite{Feher} in this paper.} $G = G_{+}G_{0}G_{-}$,
where $G_{0} = e^{\phi}$, $\phi \in\gn$, and $G_{\pm}$ are the parts of the
supergroup which corresponds to the parts $\cG_{\pm}$ of the algebra. The field
$\phi$ can be expanded as $\phi = \sum_{n} \vphi_{n}h_{n} +
\bar{\vphi}_{n}\bar{h}_{n}$, where $\vphi_{n}$ and $\bar{\vphi}_{n}$ are
bosonic superfields. The constraints can be imposed directly into the equation
of motion (since they are consistent with the WZNW dynamics). Using the
constraint equations (\ref{constr}) it is possible to show that, after a
similarity transformation with $G_{-}$, the equation of motion
$D_{-}(\hat{G}^{-1}D_{+}G)=0$ becomes
\be
	D_{+}D_{-}\phi = \{ a^{+},e^{-\phi}a^{-}e^{\phi} \}\,. \label{eqof}
\ee
This equation can be written as a zero curvature condition, i.e.
\be
	D_{-}A_{+} + D_{+}A_{-} + [ \cA_{-},\cA_{+}\} = 0\,, \label{zcr}
\ee
where $\cA_{+} = a^{+} + D_{+}\phi$, and $\cA_{-} = e^{-\phi}a^{-}e^{\phi}$.
The equation of motion (\ref{eqof}) can be derived from the
action\footnote{This action has been obtained before in ref.
\cite{Brink-Vasiliev}.}
\be
	S_{\rm Toda} = \int d^{2}zd^{2}\th str( \frac{1}{2}D_{+}\phi D_{-}\phi -
a^{+}e^{-\phi}a^{-}e^{\phi}) \,.\label{tact}
\ee
Since this action is written in $N=1$ superspace it is manifestly $N=1$
invariant, but the above action has actually $N=2$ supersymmetry, this fact can
be seen as follows. In the notations of section \ref{salg}, we write $a^{\pm} =
c^{\pm} + \bar{c}^{\pm}$, and $\phi(z^{\pm},\th^{\pm},h_{n},\bar{h}_{n})
=\Phi(z^{\pm},\th^{\pm}) + \bar{\Phi}(z^{\pm},\th^{\pm})$. Here $\Phi =
\sum_{n} \vphi_{n}h_{n}$ and $\bar{\Phi} = \sum_{n}
\bar{\vphi}_{n}\bar{h}_{n}$. We observe that $[c^{+},\bar{h}_{n}]=0$, and
similarly $[\bar{c}^{+},h_{n}] = 0$; together with the property
$str(h_{n}h_{m})=0=str(\bar{h}_{n}\bar{h}_{m})$, we see that the action can be
rewritten as
\be
	S = \int d^{2}zd^{2}\th str( D_{+}\Phi D_{-}\bar{\Phi} -
c^{+}e^{-\Phi}c^{-}e^{\Phi} -
\bar{c}^{+}e^{-\bar{\Phi}}\bar{c}^{-}e^{\bar{\Phi}})\,.
\ee
Introducing  chiral $N=2$ superfields
\bea
 \Phi_{N=2}(z^{\pm},\th^{\pm}) = \Phi_{N=1}(\tilde{z}^{\pm},\th^{\pm}) &,&
\bar{\Phi}_{N=2}(z^{\pm},\bar{\th}^{\pm}) =
\bar{\Phi}_{N=1}(\tilde{\bar{z}}^{\pm},\bar{\th}^{\pm})\,,
\eea
where $\tilde{z}^{\pm} = z^{\pm} + \th^{\pm}\bar{\th}^{\pm}$,
$\tilde{\bar{z}}^{\pm} = z^{\pm} - {\th}^{\pm}\bar{\th}^{\pm}$, the action can
be written in $N=2$ superspace as\footnote{We use
$D_{+}=\frac{\pa}{\pa\theta^{+}} + \bar{\theta}^{+}\pa_{+}$,
$\bar{D}_{+}=\frac{\pa}{\pa\bar{\theta}^{+}} + \theta^{+}\pa_{+}$, and
similarly for $D_{-}$, $\bar{D}_{-}$.}
\be
	S = \int d^{2}z\left\{ str( d^{2}\th d^{2}\bar{\th} \Phi\bar{\Phi} -  d^{2}\th
c^{+}e^{-\Phi}c^{-}e^{\Phi}  -
d^{2}\bar{\th}\bar{c}^{+}e^{-\bar{\Phi}}\bar{c}^{-}e^{\bar{\Phi}})\right\} \,,
\label{n2tda}
\ee
thereby making the $N=2$ supersymmetry manifest. The equations of motion
resulting from the above action are
\bea
	D_{+}D_{-}\Phi &=& \{\bar{c}_{+},e^{-\bar{\Phi}}\bar{c}^{-}e^{\bar{\Phi}}\}\,,
\non \\
	\bar{D}_{+}\bar{D}_{-}\bar{\Phi} &=& \{c_{+},e^{-\Phi}c^{-}e^{\Phi}\}\,.
\eea
 The equations of motion can also be obtained from a zero curvature condition
imposed on the $N=2$ Lax connection
\bea
	A_{+} =  D_{+}\Phi + \bar{c}^{+} &,& A_{-} =
e^{-\bar{\Phi}}\bar{c}^{-}e^{\bar{\Phi}} \non \,, \\
	\bar{A}_{+} = \bar{D}_{-}\bar{\Phi} + c^{+} &,& \bar{A}_{-} =
e^{-\Phi}c^{-}e^{\Phi}  \,.
\eea
It is also possible to extract the $B(n,n)$ and $B(n,n-1)$ super Toda theories
{}from the above treatment by implementing the automorphism described in section
\ref{salg}. These models have $N=1$ supersymmetry, and are described by the
action (\ref{tact}), where the fields take values in the appropriate algebra.

\setcounter{equation}{0}
\section{The General ${\cal W}$ Algebra} \label{sdb}
In this section we will derive an expression for the bracket of the
supersymmetric $\cW$ algebras associated with the generalized supersymmetric
Toda theories introduced in the previous section. Considering the great
complexity of $\cW$ algebras, it is important to be able to make general
statements of this type. The formula for the bracket encompasses the entire
class of $N=2$ $\cW$ algebras labeled by the parameter $\nu$. In particular, by
choosing the parameter $\nu$ appropriately, the formula applies to the $N=2$
supersymmetric $W_{n}$ algebras. The general spin content of the
finite-dimensional $N=2$ $W_{n}$ algebras was conjectured in ref.
\cite{Nohara,Komata}, and proven in ref. \cite{Frappat}. Supersymmetric $\cW$
algebras also appears in the supersymmetric KdV hierarchies
\cite{Huitu,Figueroa}.

We now turn to the calculation of the $\cW$ algebra bracket.
The first class constraints imposed on the currents induces a gauge invariance
of the system. The Toda field together with its equation of motion (and $\cW$
algebra) is one way to represent the gauge invariant content of the model. It
is also possible to describe the properties of the reduced system (in
particular the associated $\cW$ algebra) in other ways by supplementing the
first class constraints
\be
	\ga_{\al}=\la E_{\al},\hat{G}^{-1}D_{+}G - a^{+}\ra \approx 0 \,,
\ee
with gauge-fixing conditions. (Here $E_{\al}\in\gm\oplus\gn$.)
 The gauge-fixing conditions together with the first class constraints behave
as a set of second class constraints. At the level of the equations of motion,
the reduction can be defined on the two (chiral) currents independently,
without referring to the WZNW model. Although an $N=(2,2)$ WZNW model in
$(2,2)$ superspace is not known, an $N=(2,0)$ affine Kac-Moody algebra (in
$(2,0)$ superspace) is known \cite{Hull}. A hamiltonian reduction in $N=2$
superspace based on the $N=2$ current algebra has been developed in ref.
\cite{Ivanov}. In this reference the cases of $sl(2|1)$ and $sl(3|2)$ and their
associated $\cW$ algebras were discussed.

The Poisson bracket between the components of the WZNW current is\footnote{$X$
and $Y$ are $N=1$ (chiral) superspace coordinates, and $J=J_{+}$. In this
section we use the convention $D=\frac{\pa}{\pa\theta}+\th\pa$, $D^{2} = \pa$.
}
\be
	\{ \la J(X),T^{A}\ra,\la J(Y),T^{B}\ra \} = \la
\hat{T}^{A}[J(X),\hat{T}^{B}\}\ra\de (X-Y) - \la \hat{T}^{A},\hat{T}^{B}\ra
D_{X}\de (X-Y) \,, \label{pb}
\ee
with the definition $\de(Z_{1}-Z_{2}) = \de(z_{1} - z_{2})(\th_{1} - \th_{2})$.
The commutator of two superalgebra valued superfields $F$ and $G$ is defined as
$[F,G\} = \sum_{A,B}F_{A}G_{B}[T^{A},T^{B}\}$. Although our treatment is
classical we will occasionally use the perhaps more familiar language of
superOPE's, through the formal identification $\de(Z_{1}-Z_{2}) =
\frac{z_{12}}{\theta_{12}}$, where $z_{12} = z_{1} - z_{2} - \th_{1}\th_{2}$,
and $\th_{12} = \th_{1} -\th_{2}$. Using this rule the Poisson bracket
(\ref{pb}) can be written
\be
	\la J(Z_{1}),T^{A}\ra\la J(Z_{2}),T^{B}\ra  \sim \left[\la
\hat{T}^{A}[J(Z_{1}),\hat{T}^{B}\}\ra - \la \hat{T}^{A},\hat{T}^{B}\ra
D_{Z_{1}}\right]\frac{\th_{12}}{z_{12}} \,.
\ee

The lowest weight gauge is defined by gauge fixing $j = J-a^{+}$ to lie in
$\ker ad_{a^{-}}$. In this gauge the current can thus be written as $J=a^{+} -
W$, where $W \in \ker ad_{a^{-}}$. The total set of second class constraints
can be written as
\be
	\chi_{\al} = \la J - a^{+}, aE_{\al} \ra \approx 0\,,
\ee
where $E_{\al}$ is not lowest weight, i.e. is not annihilated by $a$. ($a$ was
defined in section \ref{salg}.) $W$ can be expanded as
\be
	W(X) = \sum_{n} W_{n+\frac{1}{2}}(X)\frac{(a^{-})^{2n}}{\la (a^{-})^{2n},
(a^{+})^{2n}\ra} + W_{n}(X)\frac{K(a^{-})^{2n-1}}{\la K(a^{-})^{2n-1},
K(a^{+})^{2n-1}\ra}\,,
\ee
with the understanding that when $d=n$, where $n$ is an integer, the above sum
only contains a finite number of terms. Recalling that $J$ is an odd field, we
see that the $W_{n}$'s are Grassmann even whereas the $W_{n+\frac{1}{2}}$'s are
Grassmann odd.
We proceed as in the bosonic case \cite{me}; the Dirac bracket is defined as
\be
	\{\cdot,\cdot\}^{*} = \{\cdot,\cdot\} - \sum_{\al\beta}\int\!\!\int dXdY
\{\cdot,\chi_{\al}(X)\}C^{-1}_{\al\beta}(X,Y)\{\chi_{\beta}(Y),\cdot\}\,.
\ee
The constraint matrix is given by $C_{\al\beta}(X,Y) =
\{\chi_{\al}(X),\chi_{\beta}(Y)\}$. Using the expression for the Poisson
bracket (\ref{pb}) together with (\ref{aprop}) we get
\be
	C_{\al\beta}(X,Y) = \la E_{\al} aU_{0}^{-1}(X) \hat{E}_{\beta}\ra\de(X-Y)\,.
\ee
Here $E_{\al}$ and $E_{\beta}$ are not lowest weight. Furthermore $U_{0}^{-1} =
1 - \cD_{0} a$, where $a$ is the (almost) inverse of $ad_{a^{+}}$, and
$\cD_{0}(Z) = D_{Z} + [W(Z),\cdot\}$. The inverse of $C_{\al\beta}$ is, as will
be shown below, given by
\be
	C^{-1}_{\al\beta}(X,Y) = -\la \hat{E}_{-\al} \hat{U}_{0}(X)A
\hat{E}_{-\beta}\ra\de(X-Y)\,, \label{cinv}
\ee
where $A=ad_{a^{+}}=[a^{+},\cdot\}$, and $E_{-\al}$ is the dual (in a suitable
normalization) of $E_{\al}$, in the sense $\la E_{-\al},E_{\ga}\ra =
\de_{\al\ga}$, a relation which defines $E_{-\al}$.
{}From this definition it follows that $E_{-\al}$ and $E_{-\beta}$ in eq.
(\ref{cinv}) are not highest weight. The above definition effectively make some
of the basis elements imaginary (because the bilinear form is non-positive
definite in general).
However, the constraints are linear in the basis element, so a rescaling of the
basis elements does not affect the end result. The $\hat{\phantom{a}}$ in
(\ref{cinv}) acts in the following way on $\cD_{0}a$, $\cD_{0}(Z)a \rightarrow
-(D_{Z} + [\hat{W}(Z),\cdot\})a$. For future reference we notice that the
following closure relations hold
\bea
	\sum_{\al} \hat{E}_{\al}\ra\la \hat{E}_{-\al} =1 \,,\non \\
	\sum_{\al} \hat{E}_{-\al}\ra\la E_{\al} =1  \,. \label{cmplt}
\eea
We want to show
\be
	\sum_{\beta}\int DY C_{\al\beta}(X,Y)C^{-1}_{\beta\ga}(Y,Z) =
\de_{\al\ga}\de(X-Z)\,. \label{cc}
\ee
In order to show this, i.e
\be
-\sum_{\beta}\int DY  \la E_{\al} aU_{0}^{-1}(X) \hat{E}_{\beta}\ra\de(X-Y) \la
\hat{E}_{-\beta} \hat{U}_{0}(Y)A \hat{E}_{-\ga}\ra\de(Y-Z) =
\de_{\al\ga}\de(X-Z)\,,
\ee
we use the following property
\be
	\int DY \de(X-Y)f(Y) = -f(X) \,.\label{sint}
\ee
We also use the fact that when moving $U$ through $DY$ it acquires a
$\hat{\phantom{a}}$. Equation (\ref{cc}) now follows using the closure
relation, together with $\la E_{\al},\hat{E}_{-\ga}\ra = \la
E_{-\ga},E_{\al}\ra = \de_{\al\ga}$. We can now write down the expression for
the Dirac bracket between the components of the gauge-fixed current (the first
term in the general expression for the Dirac bracket has been included)
\bea
	&& \{\la J(Z_{1}),T^{A}\ra,\la J(Z_{2}),T^{B}\ra\}^{*} = - \la
\hat{T}^{A}\cD_{0}(Z_{1}) \hat{T}^{B}\ra\de(Z_{1}-Z_{2}) - \non \\ && -
\sum_{\al\beta}\int \!\! DX\!\!\int \!\!DY \{\la
J(Z_{1}),T^{A}\ra,\chi_{\al}(X)\}C^{-1}_{\al\beta}(X,Y)\{\chi_{\beta}(Y),\la
J(Z_{2}),T^{B}\ra\} \label{undb}\,.
\eea
Calculating the various factors using the definition of the second class
constraints we can write the second term in (\ref{undb}) as
\bea
	&& -\sum_{\al\beta}\!\int\!\!\int\!\! DX DY \la
\hat{T}^{A}\cD_{0}(Z_{1})a\hat{E}_{\al}\ra\de(Z_{1}-X)\cdot \non \\
&&\phantom{AAAAAAAA}
 \cdot\la \hat{E}_{-\al} \hat{U}_{0}(X)A \hat{E}_{-\beta}\ra\de(X-Y)\la
E_{\beta}a\cD_{0}(Y)\hat{T}^{B}\ra\de(Y-Z_{2}).
\eea
Using the closure relations (\ref{cmplt}) together with (\ref{sint}), we get
\be
	-\la \hat{T}^{A}\cD_{0}(Z_{1}) aU_{0}(Z_{1})\cD_{0}(Z_{1})
\hat{T}^{B}\ra\de(Z_{1}-Z_{2})\,.
\ee
We finally arrive at the following expression for the Dirac bracket in the
lowest weight gauge
\be
	\{ W_{A}(X),W_{B}(Y) \}^{*} = - \la \hat{T}^{A} U_{0}(X)\cD_{0} (X)
\hat{T}^{B} \ra\de(X-Y) \,, \label{db}
\ee
with $U=(1-{\cal D}_{0}a)^{-1}$, and $W_{A} = \la W,T^{A}\ra$.
We have thus constructed the bracket of the ${\cal W}$ algebra in the
realization where the generators are linear in the current. It is a simple
matter to (formally) transform to the language of SOPE's
\be
	 W_{A}(Z_{1})W_{B}(Z_{2}) = -\la \hat{T}^{A} U_{0}(Z_{1})\cD_{0} (Z_{1})
\hat{T}^{B} \ra\frac{\theta_{12}}{z_{12}} \,. \label{dbope}
\ee
As an example we calculate the OPE between the superspin 1 generator $W_{1}$,
the super \emt $W_{\frac{3}{2}}$, and the other $\cW$ algebra generators. The
method used to derive the results (\ref{sope}) is simple: One starts by writing
$W_{A}(Z_{1})W_{B}(Z_{2}) \sim \la
\hat{T}^{A}\sum_{n}(a\cD_{0}(Z_{1}))^{n}\cD_{0}(Z_{1})\hat{T}^{B}\ra$. The next
step is to let the operators $a$ and $\cD_{0}$ act on $\hat{T}^{B}$, and use
the properties of the bilinear form (\ref{biprop}) including the property
(\ref{aprop}). Since $a$ decreases the grade and $\cD_{0}$ never increases the
grade only a finite number of terms in the sum above will contribute. The
result is as expected
\bea
	W_{1}(Z_{1})W_{n}(Z_{2}) &\sim& \frac{\hat{c}}{2z^{2}_{12}}\de_{s,1} +
2n\frac{\theta_{12}}{z_{12}}W_{\frac{2n+1}{2}}(Z_{2})  \,,\non \\
	W_{1}(Z_{1})W_{\frac{2n+1}{2}}(Z_{2}) &\sim&
-n\frac{\theta_{12}}{z^{2}_{12}}W_{n}(Z_{2}) -
\frac{D_{2}W_{n}(Z_{2})}{2z_{12}} - \frac{1}{2}
\frac{\theta_{12}}{z_{12}}\pa_{2}W_{n}(Z_{2}) \,,  \non \\
	W_{\frac{3}{2}}(Z_{1})W_{s}(Z_{2}) &\sim&
\frac{\hat{c}}{4z^{3}_{12}}\de_{s,\frac{3}{2}} +
s\frac{\theta_{12}}{z^{2}_{12}}W_{s}(Z_{2}) + \frac{D_{2}W_{s}(Z_{2})}{2z_{12}}
+ \frac{\theta_{12}}{z_{12}}\pa_{2}W_{s}(Z_{2})\,, \label{sope}
\eea
where $\hat{c}=2d(d-1)$. The central charge in the Virasoro algebra is given by
$c=\frac{3}{2}\hat{c}=3d(d-1)$.
When $d=2$ only two generators survive, namely $W_{1}$ and $W_{\frac{3}{2}}$.
$W_{1}$ contains one field of spin 1 and one of spin $\frac{3}{2}$;
$W_{\frac{3}{2}}$ contains one field of spin $\frac{3}{2}$ and one of spin 2
(the \emt). Together these generators span the well known $N=2$ super conformal
algebra \cite{Ademollo1,Ademollo2}. When $d=n$ we obtain instead the $N=2$
supersymmetric $W_{n}$ algebra.
The general expression for the central charges in the algebra can also be given
as
\be
W_{s}(Z_{1})W_{s'}(Z_{2}) \sim \frac{c_{s}}{z^{2s}_{12}}\de_{s,s'} + \ldots
\ee
where
\bea
	c_{\frac{2n+1}{2}} &=& \frac{n}{2}c_{n} \non \\
	c_{n} &=&
\frac{(-1)^{n+1}}{2^{n-1}}\frac{(n-1)!}{(2n-1)!!}\prod_{l=-n}^{l=n-1}(d+l)\,.
\eea
We close this section by noting that it would be interesting if the $\cW$
algebra bracket could be written in $N=2$ superspace language, thereby making
the $N=2$ supersymmetry manifest, this would entail rewriting the formula
(\ref{db}) in $N=2$ superspace.
\setcounter{equation}{0}
\sect{The Miura Transformation}
In this section we will describe a free field realization of the super $\cW$
algebra introduced in the previous section. We will follow the treatment of the
bosonic case \cite{me}. The goal is to derive a closed expression for the $\cW$
algebra generators expressed in terms of free superfields. We start by
partially fixing the gauge in the following way
\be
	J \approx a^{+} - D_{+}\Ups - W \label{fix}
\ee
where $\Ups\in\cG_{0}$. The superfield $W\in \ker ad_{a^{-}}$ in (\ref{fix}) is
in general different from the $W$ used in section 4, however when we constrain
$\Ups$ to zero by further gauge fixing, the two $W$'s become equal. If we
instead gauge-fix $W$ in (\ref{fix}) to be zero, we obtain the so called
diagonal gauge. This gauge is only accessible locally; but nevertheless it is
extremely useful, since it gives rise to a free-field realization of the $\cW$
algebra, since the components of $\Ups$ satisfy $D_{+}D_{-}\Ups_{A} = 0$, as
well as free field (Dirac) superbrackets.
The gauge fixing (\ref{fix}) imply that the remaining unfixed first class
constraints are of the form
\be
\vrho_{\al} = \la J-a^{+}, ah_{\al}\ra \approx 0\,, \label{res}
\ee
where the $h_{\al}$'s span a basis in the subspace of grade zero elements.
The $\cW$ algebra is spanned by gauge invariant combinations of $J$'s. Under a
gauge transformation generated by the residual first class constraints
(\ref{res}) the $\cW$ algebra generators are invariant i.e. $\de {\rm\sf W}[J]
= \{\vrho(X),{\rm\sf W}[J]\}^{*}\approx 0$. Here ${\rm\sf W}[J]$ is an
arbitrary $\cW$ algebra generator. On the constraint surface we have ${\rm\sf
W}[J]={\rm\sf W}[D_{+}\Ups,W]$, since $J$ is then of the form (\ref{fix}). For
a constraint of the form  $\vrho(X) = \la J(X)-a^{+},a\xi\ra$ we get
\be
	\de  {\rm\sf W}= \left[ \int \{\vrho(X),W_{A}(Y)\}^{*}dY\frac{\de}{\de
W_{A}(Y)} + \int\{\vrho(X),\pap \Ups_{A}(Y)\}^{*}dY\frac{\de}{\de D_{+}
\Ups_{A}(Y)}\right] {\rm\sf W}\approx 0\,. \label{wprop}
\ee
Here we have used the definition $\frac{\de}{\de W_{A}(X)}W_{B}(Y) =
\de_{AB}\de(X-Y)$, and similarly for $\frac{\de}{\de D_{+}\Ups_{i}(X)}$
 The next step is to calculate the Dirac brackets appearing in (\ref{wprop}).
We start with $\{\vrho(X),D_{+} \Ups(Y)\}^{*}$. As in the bosonic case
\cite{me} it can be proven that only the first term in the general expression
for the Dirac bracket contributes, hence
\be
	 \{\vrho(X),\la D_{+} \Ups(Y),h \ra\}^{*} =  \{\vrho(X),\la
D_{+}\Ups(Y),h\ra\}= \la \xi,h\ra\de(X-Y) \,.
\ee
The other Dirac bracket in (\ref{wprop}) can be shown to equal
\bea
	\{\vrho(X),\la W(Y),T^{A}\ra \}^{*} &\hspace{-2mm}=&  -\la \xi a\cD(X)
\hat{T}^{A}\ra\de(X-Y) - \la \xi a\cD(X)a U(X) \cD(X)
\hat{T}^{A}\ra\de(X-Y)\non   \\& \hspace{-2mm}=& -\la
\xi,V(X)\hat{T}^{A}\ra\de(X-Y)\,, \label{lwdb}
\eea
where $U(X) = (1- \cD(X)a)^{-1}$, $\cD(X) = D_{X} + [D\Ups(X),\cdot\} +
[W(X),\cdot\}$, and $V(X) =(1- a\cD(X))^{-1}$. The proof of equation
(\ref{lwdb}) follows closely the calculation performed in the bosonic case
\cite{me}. We can move the $\hat{\phantom{a}}$ in (\ref{lwdb}) from $T^{A}$ to
$V(X)$, using the properties of the trace. Inserting the above results into
(\ref{wprop}) and performing the integration over $Y$, we obtain the equation
\be
	\left[\la \xi,\frac{\de}{\de D_{+}\Ups}\ra - \la
\xi,V(D_{+}\Ups,W)\frac{\de}{\de W}\ra\right]{\rm\sf W}[D_{+}\Ups,W] = 0\,,
\label{gov}
\ee
where we have used the notation $\frac{\de}{\de W} =
\sum_{A}T^{A}\frac{\de}{\de W_{A}}$. In this formula the $T^{A}$'s are highest
weight elements satisfying $W_{A} = \la W,T^{A} \ra$, which implies $\la \eta,
\frac{\de}{\de W(X)}\ra W(Y) = \eta\de(X-Y)$. The integrability condition of
the variational equation (\ref{gov}) is satisfied as a consequence of the fact
that the Dirac bracket satisfies the (super)Jacobi identity.
It is possible to show that the solution to equation (\ref{gov}) is
\be
	{\rm\sf W}[D_{+}\Ups,W] = \exp(-\int_{0}^{1}ds\int dY\la
D_{+}\Ups(Y),V[sD_{+}\Ups(Y),W(Y)]\frac{\de}{\de W(Y)}\ra){\rm\sf W}[W]\,.
\ee
It is possible, using the above formula, to derive expressions for the $\cW$
algebra generators expressed in terms of the free superfield $\Ups$, using
${\rm\sf W}[D_{+}\Ups] = {\rm\sf W}[D_{+}\Ups,W]|_{W=0}$, i.e. going to the
diagonal gauge.
As an example we give the first few generators. We get
\bea
	W_{1}[D_{+}\Ups] &=& \frac{1}{2}\la D_{+}\Ups r D_{+}\Ups\ra + \la \frac{(K +
2\nu)}{2} D_{+}^{2}\Ups\ra \,, \non \\
	W_{\frac{3}{2}}[D_{+}\Ups] &=&  -\frac{1}{2}\la D_{+}\Ups D^{2}_{+}\Ups\ra +
\la T^{0}D^{3}_{+}\Ups \ra\,, \label{freegen}
\eea
where the operator $r$ has been defined in section \ref{salg}. To obtain the
first expression we made use of the fact that $K=r$ when acting on ${\bf
Z}_{2}$ odd elements, and furthermore that $rA=Ar$ when acting on neutral
elements. From (\ref{freegen}) we see that the existence of $r$ is crucial for
the existence of the spin 1 generator.
When $d=2$ the above expressions constitute a free field realization of the
$N=2$ super conformal algebra. Using the same method as in the bosonic case it
is possible to prove that ${\rm\sf W}[D_{+}\Ups] = {\rm\sf W}[-D_{+}\phi]$,
i.e. that the $\cW$ algebra generators are the same\footnote{The relative sign
is a result of our conventions.} when expressed in terms of the free fields and
the Toda fields. E.g. we see that the superspin $\frac{3}{2}$ generator above
is precisely the (improved) \emt of the Toda theory described in section
\ref{swznw}. It is of course important to study the quantization of the general
$\cW$ algebra. We will return to this problem in a future publication.

\bigskip\bigskip\noindent{\Large\bf Acknowledgement}

\bigskip\noindent The author wishes to thank L. Brink and M. Vasiliev for
stimulating discussions.

\end{document}